\documentclass[epj]{svjour}

\newcommand{\beq}{\begin{equation}}
\newcommand{\eeq}{\end{equation}}
\newcommand{\beqa}{\begin{eqnarray}}
\newcommand{\eeqa}{\end{eqnarray}}
\newcommand{\tr}{{\rm Tr}}
\usepackage{graphicx,amssymb,mathrsfs}
\usepackage{fancyhdr}
\def\makeheadbox{{
 }}
\def\mytitle{My title} 
\def\myauthors{My name}  
\def\mytype{My type of session}
\def\mysession{My session}

\def\mytitle{Local SUSY-breaking minima in $N_f=N_c$ SQCD?} 
\def\myauthors{Andrey Katz}    
\def\mytype{Contributed Talk}    
\def\mysession{Theoretical Models}

\begin{document}
\title{Local SUSY-breaking minima in $N_f=N_c$ SQCD?}
\author{Andrey Katz 
\thanks{\emph{Email:} andrey@physics.technion.ac.il}%
}                     
%
%
\institute{Technion - Israel Institute of Technology}
%
\date{}
\abstract{
We study non-supersymmetric minima in $N_f=N_c$ SQCD conjectured by
Intriligator, Seiberg and Shih (ISS).  We show that the existence of such minima depends
on the signs of three non-calculable parameters and that no evidence can be inferred
by deforming the theory.  We illustrate this by demonstrating that the conjectured minimum 
is destabilized in a different deformation of $N_f=N_c$ SQCD. We also comment briefly 
on the phenomenological consequences of this instability. 
} 
\maketitle
\section{Introduction}
\label{intro}
It was shown by Intriligator, Seiberg and Shih (ISS)~\cite{Intriligator:2006dd} that supersymmetry breaking 
in meta-stable vacua is a common phenomenon which occurs in a wide range of models. In particular they showed that 
such vacua are present in massive supersymmetric QCD (SQCD) with the number of colors $N_c$ and number of 
flavors $N_f$ in the range $N_c<N_f<\frac23 N_c$. It was 
demonstrated that the supersymmetry breaking minimum is located near the origin of field space, far from the
supersymmetric minimum, such that the lifetime of the supersymmetry breaking minimum is sufficiently long. The
analysis was done using the Seiberg dual theory~\cite{Seiberg:1994pq}, which is weakly coupled.

ISS also considered the case  $N_f=N_c$. This theory confines at low
energies and the quantum moduli space is deformed. 
Since the deformation of the moduli space does not allow all the fields to be near the origin of 
the field space calculability is lost. 
In order to make further progress ISS deformed this theory adding one more massive flavor
(which restores calculability) and then decoupling it, taking the mass of the additional flavor to be 
infinitely large. By using this deformation ISS conjectured that the local minimum also exists 
in $N_f=N_c$ SQCD.

In our current work~\cite{Katz:2007gv} we revisit the case $N_f=N_c$. In the first part of my talk
I will briefly review the main results of ISS. It will be also shown that the existence of the meta-stable vacuum 
in the massive $N_f=N_c$ SQCD
is solely dependent on the signs of three non-calculable terms in the K\"ahler potential and hence no information
about this minimum can be gained by deforming the theory. To illustrate this point I will introduce another
deformation of $N_f=N_c$ SQCD. The potential of the proposed deformation possesses an extremum similar to ISS,
but it is a saddle point rather than a minimum. I will also comment on the phenomenological
consequences of this instability, concentrating on the ``Pentagon" model~\cite{Banks:2006ma}.
         
\section{Review of the ISS model, $N_f=N_c$ conjecture}
\label{sec:1} 
Consider SQCD with $N_c+1\le N_f <\frac32 N_c$ and the tree level superpotential
\beq
W=(m_Q)_{ij}Q_i \bar Q_j
\eeq
where $Q$ and $\bar Q$ denote quarks and antiquarks respectively. Assume also that the rank of the matrix $m_Q$
is larger than the number of colors in the magnetic dual theory $N_f-N_c$. Clearly, this model possesses a 
supersymmetric global vacuum. Nonetheless, as was shown in~\cite{Intriligator:2006dd}, the theory has a local
non-supersymmetric vacuum near the origin of the moduli space. Supersymmetry is broken by a rank condition in 
the magnetic dual theory. One of the most crucial points for this analysis was that the supersymmetric 
quark masses $(m_Q)_{ij}$ are much lower than the scale of the magnetic theory $\Lambda$. The smallness of the 
parameter $m_Q/\Lambda$ is important for two reasons. It enables us to increase the distance between 
the supersymmetric and non-supersymmetric vacua in field space, rendering the lifetime of the meta-stable
solution sufficiently long. But it is also important in order to ensure the calculability of the model, 
ensuring that the contributions of non-calculable terms in K\"ahler potential are neglegible.    

ISS also considered the case  $N_f=N_c$. One can think about this theory as the limit of $N_f=N_c+1$ SQCD when
the mass of one of the flavors is taken to be infinitely large. Although the theory with 
$N_f=N_c+1$ confines, it has the same calculable supersymmetry 
breaking meta-stable vacuum as the theories in the free magnetic phase. Naively, one could trace the 
behavior of this minimum when the limit $N_f=N_c$ is taken, and come to the conclusion that the theory with 
$N_f=N_c$ has the same minimum. Nevertheless, $N_f=N_c $ SQCD is very different from the cases
with with larger number of flavors: even if we demand $m_Q\ll \Lambda$, the contribution of the noncalculable 
K\"ahler potential is not negligible, moreover it dominates the calculable contributions.  

When $N_f=N_c$ SQCD confines the quadratic approximation for the K\"ahler potential fails and one should 
take into account higher order terms. 
Consider the K\"ahler metric at next to leading order: using the global 
symmetries of the theory it can be parametrized as follow\footnote{Hereafter $B_+$ and $B_-$ denote the heavy 
and the light combination of baryons respectively} :
\beq\label{metrica}
g^{-1}_{M^\dagger M}\sim \alpha \frac{ \tr(M^\dagger M)}{\Lambda^2}+\beta \frac{\tr M \tr M^\dagger}{\Lambda^2}
+\gamma \frac{(B_++B_+^\dagger)^2}{\Lambda^2}+\cdots
\eeq   
Here we used the fact that the baryon symmetry is spontaneously broken. The solution, conjectured by ISS, is 
located along the baryonic branch and  ${\rm Im}(B_+)$ is supposed to be a Goldstone boson. The odd powers 
of ${\rm Re}(B_+)$ are forbidden by an unbroken subgroup 
${\mathbb Z}_{2N}$ of the anomalous axial symmetry $U(1)_A$. The coefficients $\alpha,\ \beta $ and $\gamma$ 
are non-calculable and probably of order one; neither their precise numerical values nor their signs are known.

Assume for simplicity that the quark mass matrix is proportional to the identity.  Then the non-vanishing 
F-term in the conjectured local minimum is $F_{\tr M}=m_Q\Lambda$. We get the following contribution 
to the scalar potential from the non-calculable terms in K\"ahler metric:
\beq\label{ncpoten}
V\sim m_Q^2 \left( \alpha \ \tr(M^\dagger M) +\beta \ \tr M\tr M^\dagger +\gamma \ (B_++B_+)^2 \right)
\eeq     
Namely we get non-calculable contributions to the squared masses of baryons and mesons of order 
$m_Q^2$.  Note that these contributions are of the same order of magnitude both in $N_f=N_c$ and
$N_f=N_c+1$ cases. Now let us compare these contributions to the calculable contributions to the masses 
in the meta-stable vacuum. 

Consider the theory with $N_f=N_c+1$. Denote its confining scale by $\hat \Lambda$ to distinguish it 
from $\Lambda$, the confining scale of $N_f=N_c$. It was shown in~\cite{Intriligator:2006dd} that most of the fields
get squared masses of order ${\cal O}(m_Q \hat \Lambda)$. There are also pseudo-moduli, which are stabilized 
at the one-loop level and get squared masses of order 
${\cal O}\left(\frac {m_Q \hat \Lambda}{16\pi^2}\right)$. Clearly 
these contributions are larger than those of~(\ref{ncpoten}).   

Now assume that we take the mass of one of the flavors (denote this mass by $(m_Q)_{N_c+1}$)
to be large, leaving all other
masses of the same order of magnitude $m_Q$ and small. The tree level masses are now of order
\beq\label{calcmass}
m^2\sim {\cal O}\left(\frac{m_Q^2\hat \Lambda}{(m_Q)_{N_c+1}} \right)~,
\eeq
and the masses of pseudo-moduli are suppressed by an extra factor of $16\pi^2$. 
Taking the limit $(m_Q)_{N_c+1}\to \infty$
we get the theory with $N_f=N_c$ and the confining scale
\beq
\Lambda^{2N_c}=(m_Q)_{N_c+1}\hat \Lambda^{2N_c-1}
\eeq
Clearly, taking this limit we should also take the scale $\hat \Lambda$ to zero, preserving the confining scale
of $N_f=N_c$ finite. In this limit the 
expression~(\ref{calcmass}) tends to zero, but the noncalculable contributions are supposed to stay finite.
Namely when we get to the limit $N_f=N_c$ the existence of the minimum
depends only on the signs of $\alpha,\ \beta, \ \gamma$.  
\section{Another deformation}
\label{sec:2}

In this section we introduce another deformation of massive $N_f=N_c$ SQCD. We will try to find in
this deformation the same extremum, as was found by ISS, and to check whether it is a minimum.

Consider SQCD with $N_f=N_c$ and singlets $S_{ij}$, $T$ and $\bar T$. Let $S$ be in 
the $({\bf N}, \bar {\bf N})$ 
representation of the global flavor symmetry $SU(N_f)_{L}\times SU(N_f)_{R}$\footnote{Clearly with the massive 
quarks this symmetry is broken, but since the breaking is small we will still use these notations.} 
while $T$ and $\bar T$ are singlets of the flavor symmetry. We couple these singlets to the quarks and add the 
singlet masses. The tree level superpotential is 
\beqa \nonumber
\Delta W &= \lambda S_{ij} Q_i \bar Q_j + &(m_Q)_{ij} Q_i \bar Q_j +\kappa (T\det Q +\bar T\det \bar Q)+ \\ 
&&\frac{m_S}{2} S^2 +\frac{m_T}{2}(T^2+\bar T^2)~.
\eeqa   
This model is nothing but the Intriligator-Thomas-Izawa-Yanagida 
model~\cite{Intriligator:1996pu},\cite{Izawa:1996pk} with 
massive singlets\footnote{Originally this model with massless singlets was constructed as an example of 
the model which breaks 
supersymmetry dynamically, but this is irrelevant for our further discussion.}. The singlet masses are necessary
if we expect to observe any effect, proportional to the quark mass; without these masses $m_Q$ can be absorbed
into redefinitions of the singlets $S_{ij}$.

Hereafter we assume for simplicity that the quark mass matrix is proportional to the identity. The full low energy
superpotential is 
\beqa\nonumber
W & = & {\cal A}(\det M - B \bar B -\Lambda^{2N})+
 \lambda \tr (MS) +\kappa (BT +\bar B \bar T)\\ \label{superpoten}
&&m_Q \tr M +\frac{m_S}{2} S^2 +\frac{m_T}{2}(T^2+\bar T^2)~. 
\eeqa
Since we are interested in vacua of the SQCD with $N_f=N_c$ without any singlets we will study our theory
in the limit that the gauge singlets are almost decoupled from the other fields. Obviously we can not completely 
decouple the singlets from the theory since then calculability is lost. Hence we will carry out
our calculation in the limit where the couplings are finite and small. At the end of this analysis we will revisit
the issue of calculability and determine the precise range of validity for our calculations. There are
two equivalent ways to decouple each singlet from the theory: one can either take its mass to be infinitely 
large or take its coupling to the mesons and baryons to be small. Appropriately rescaling the fields one can 
see, that these two ways are equivalent: one should take $\lambda/m_S^2 \to 0$ to decouple 
the singlets $S_{ij}$ and $\kappa^2/m_T^2 \to 0$ to decouple $T$ and $\bar T$.   

Now we look for the extrema of the potential at the decoupling limit. We expect to find both supersymmetric 
and non-supersymmetric solutions. The supersymmetric solution should stay at a finite distance from the origin 
once the decoupling limit is taken. In order to match the non-supersymmetric ISS solution 
the meson should be stabilized at the 
origin of fields space. Moreover, we expect that the vacuum energy of the non-supersymmetric solution 
scales as $V\sim m_Q^2$. 

Assuming the superpotential~(\ref{superpoten}) one can find three different supersymmetric solutions. In two of 
these solutions the meson scales as $M\sim \left(\frac{m_T}{\kappa^{2}}\right)^\frac{1}{N-1}$. Since the distance 
between these solutions and the origin is infinite at the decoupling limit we will not consider these solutions 
further. Only the third solution is relevant for our analysis : $M\sim \Lambda$ with
all other fields being stabilized at zero in the decoupling limit. This solutions lies far enough from the 
origin, so if one succeeds to find a SUSY breaking minimum near the origin, it could be made meta-stable.

One also finds a non-supersymmetric extremum. Treating both $B_+$ and $B_-$
as dynamical fields\footnote{Strictly speaking 
$B_-$ should not be treated as dynamical. It is heavy and should be integrated out of the theory, but as we 
show in~\cite{Katz:2007gv} such proper treatment does not change  qualitatively any of our results.} 
and 
dropping all numerical factors of order one we find
\beq
M\sim \left(\frac{\lambda^2}{m_s} \frac{m_T}{\kappa^2}\right)^\frac{1}{N-2} \Lambda,  \ \ B_- \sim \Lambda
\eeq
All other fields are stabilized near the origin in the decoupling limit. 
To recover the ISS solution we should decouple the singlets $S_{ij}$ faster than $T$ and $\bar T$:
\beq
\frac {{\lambda^2}/{m_S}}{{\kappa^2}/{m_T}} \to 0~.
\eeq

In order to understand whether this solution is a stable minimum, we study the mass spectrum of this solution. 
All the fields $B$ and $T$ get positive squared masses, their spectrum is supersymmetric and decoupled from the rest
of the fields.
The fermionic spectrum of the fields $\tr M$ and $\tr S$ contains exactly two  massless states 
(Goldstino) and two massive state with  mass of order $m_S^2+(\lambda\Lambda)^2$. In the ISS model the mesino was
the Goldstino. Here the  
Goldstino is a superposition of the mesino
and the singlet-fermion  and as expected, it reduces to the mesino in the decoupling limit. 
The spectrum of this sector is not
supersymmetric and there is only one SUSY-breaking entry to the mass matrix. Hereafter we call this 
entry $\xi$. The full expression for $\xi$ is rather cumbersome, but for large $N$ it approximately behaves as
\beq
\xi\sim \frac{\lambda^2\Lambda^2}{m_S} m_Q^*~.
\eeq
As expected, it is proportional to $m_Q$, since the quark mass was supposed to be responsible for 
triggering of SUSY-breaking.  The degeneracy between the scalar states is removed: the two states 
which are degenerate in the fermionic case are now split around the fermion mass and the splitting is
proportional to $\xi$. But two of the four fermionic states were precisely massless! Therefore we get
one tachyonic mode. Thus the point, which tends to the ISS-conjectured
minimum is a saddle point in the ITIY deformation, with precisely one unstable
direction. One could still hope that this tachyonic direction may be lifted by Coleman-Weinberg potential.
Unfortunately it does not happen since $V_{{\rm CW}}\propto \xi$ and loop-suppressed, thus this contribution
to the effective potential is smaller than a tree-level. 

We can now estimate the range of validity of our calculations. In order to trust our calculations we should demand
that the noncalculable contributions are much smaller than the calculable ones. The mass squared of the lowest
(tachyonic) mode is 
roughly $|\xi|$ and as we have already mentioned the non-calculable contribution is always $m_Q^2$.  Namely our 
calculations are valid in the range where (for large $N$)
\beq
\frac{(\lambda \Lambda)^2}{m_S} \gg m_Q
\eeq
To maintain this condition one can choose for example $m_Q\ll \lambda \Lambda \ll m_S \lesssim \Lambda$. Note that 
$m_Q$ should be the smallest mass parameter of the theory. Namely we can not take a "true" decoupling 
limit with $\lambda \Lambda$  lower than $m_Q$ without loosing calculability. 

\begin{figure}
\includegraphics[width=0.35\textwidth,height=0.25\textwidth,angle=0]{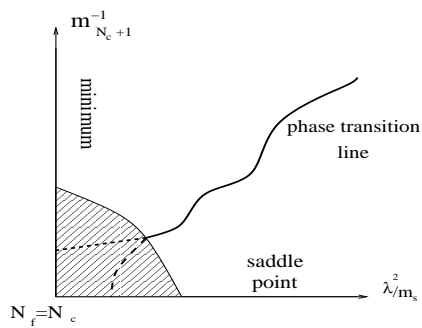}
\caption{Two deformations of $N_f=N_c$ SQCD. In the shaded region calculability is lost.}
\label{phasetrans}       
\end{figure}

This situation is summarized in figure~\ref{phasetrans}. Consider the theory with $N_f=N_c+1$ coupled 
to the massive singlets of ITIY theory. Clearly, when the couplings of quarks to the singlets are weak 
and the masses of the quarks are small the theory has a meta-stable minimum near the origin. 
On the other hand if the couplings to the
singlets are significant and the mass of one of the flavors is very high we find a saddle point near the 
origin; the ISS extremum undergoes a phase transition.  But note that if both $m_{N_c+1}^{-1}$ and
 $\frac{\lambda^2}{m_S}$ are small calculability is lost and we do not know how the phase transition 
line behaves. Hence it is impossible to determine if the extremum is a minimum or a saddle point in the 
$N_f=N_c$ SQCD.

\section{Phenomenological consequences.} 
\label{sec:3} 
The instability described in the previous section, has direct consequences on the ``Pentagon" model, which 
I will briefly discuss in this part of the talk.

First I introduce the ``Pentagon", as it appears in the ``remodelled" version~\cite{Banks:2006ma}.
Consider massive SQCD with $N_f=N_c=5$, where the mass matrix is proportional
to the identity  in the flavor basis. This mass term breaks the flavor symmetry of the model down to 
$SU(5)_{diag}$. The Standard Model (SM) is embedded into this weakly gauged flavor $SU(5)_{diag}$.
The model assumes that the local SUSY-breaking minimum, conjectured by ISS, exists near the origin of field 
space. SUSY is broken in this local minimum and the breaking is mediated to the SM fields by the off-diagonal 
components of the meson. In order to avoid fine-tuning and get viable phenomenology one should relate 
the $\mu$-term to the superpartner scale.    
To get a $\mu$-term of the desired magnitude the singlet $S$ is introduced, which 
couples both to the Higgses of the SM and the quarks of the hidden sector. When this singlet develops a VEV of 
the right size, it produces a required $\mu$-term.  $S$ couples to the quarks through the 
hypercharge in order not to break the SM subgroup of the $SU(5)$ and the explicit $\mu$-term is forbidden by 
some discrete symmetry. The relevant part of the superpotential is:
\beq
W=m_Q\tr M +SH_u H_d + \lambda S\tr(YM)~.
\eeq     
Here $Y$ denotes the hypercharge generator \\ $Y = {\rm diag} (1, 1, 1, -\frac32, -\frac32)$. 

It is clear from the previous discussion that one can not be sure that any meta-stable vacuum exists in the given 
setup. But at this stage let us believe the ISS conjecture and assume that the required minimum exists. 
First assume  $m_Q\ll \Lambda_5$ (which is necessary for the conjectured vacuum to be meta-stable)\footnote{Hereafter
$\Lambda_5$ denotes the "Pentagon" confining scale} and $\lambda\ll 1$.
 
Our statement is that with these assumptions one can not get a viable spectrum for the superpartner masses from the ``Pentagon" model. Recall that all the bosons will get noncalculable  masses squared 
of order $m_Q \Lambda_5$. The masses of the fermions are much smaller since they involve some power 
of $\lambda$, and we have assumed that it is small. Namely we generate a large positive messenger supertrace. 
But as it 
was shown in~\cite{Poppitz:1996xw} such a supertrace will give large \emph{negative} contributions to the squark 
masses squared.

To avoid this phenomenological disaster one should take $\lambda$ to be sufficiently large. But taking it to be large
$S$ acts precisely as $\tr S$ in our previous discussion. Calculability is restored but the conjectured minimum
is destabilized. Thus the ``Pentagon" is excluded when $\lambda $ is either too large or too small. 
In the intermediate range the model is completely non-calculable and one can make no statement about the model. 
Although it is quite possible that at this range of energies one gets very complicated potential with more
local minima,
it is also worth to notice, that this conjectured minimum
can not be directly related to the ISS minimum. Since $\lambda $ and $m_Q$ should be both relatively large, one 
can not rely on the expansion~(\ref{metrica}) and we should know the entire K\"ahler potential in order to 
understand the nature of the conjectured minimum near the origin. Consequently, if the conjectured 
local minimum of the Pentagon exists, one can not rely on~\cite{Intriligator:2006dd}, 
making any statement about its metastability
or the properties of the related sparticle spectrum.      
 
\section{Conclusions}
\label{sec:4}

We conclude that there is no clear indication whether the meta-stable SUSY breaking minimum 
exists in $N_f=N_c$ SQCD. We showed that no additional evidence can be gained by deforming the theory. A point
which was a minimum of one deformation is a saddle point of another deformation. We have also demonstrated that 
the instability which we found significantly restricts the range, for which the ``Pentagon" model is not excluded. 
In particular we emphasize that in this range the ``Pentagon" is necessarily non-calculable and its
phenomenologically viable minimum, even if it 
exists, can not be directly related to the ISS-conjectured minimum.
 
\vspace{0.5cm}
\emph{Acknowledgments}: I thank Yael Shadmi and Tomer Volansky for collaboration and also for critical reading of this 
note and their useful comments. 
I also thank the organizers of SUSY07 for 
giving me an opportunity to present our work.

%

%

%
%

\end{document}